\documentclass[letterpaper,12pt]{article}

\usepackage{a4wide}

\usepackage{mathptmx}
\usepackage[scaled=.90]{helvet}
\usepackage{courier}

\usepackage{amsmath}
\usepackage{eucal}

\usepackage{graphicx}
\usepackage{url}
\usepackage{xspace}
\usepackage{ifthen}

\usepackage{apacite}

\newcommand{\Tool}[1]{\textsc{#1}\xspace}
\newcommand{\Address}[1]{$^{\rm #1}$}
\newcommand{\cf}{\textit{cf}\xspace}
\newcommand{\eg}{e.g.\xspace}
\newcommand{\ie}{i.e.\xspace}
\newcommand{\vs}{\textit{vs}\xspace}
\newcommand{\wrt}{with respect to\xspace}
\newcommand{\Kmer}{$k$-mer\xspace}
\newcommand{\Kmers}{$k$-mers\xspace}

\newcommand{\Opt}[1]{\texttt{#1}}

\newcommand{\LW}{$<\!L,W\!>$\xspace}
\newcommand{\LWp}[2]{$L=#1$, $W=#2$\xspace}

\newcommand{\Paragraph}[1]{\paragraph{#1}}

\newcommand{\Balibase}{BAliBASE\xspace}

\newcommand{\Clustalw}{\Tool{ClustalW}}
\newcommand{\Dialign}{\Tool{Dialign}}

\newcommand{\Muscle}{\Tool{MUSCLE}}

\newcommand{\Phylip}{\Tool{PHYLIP}}
\newcommand{\Phylogen}{\Tool{PhyloGen}}
\newcommand{\Probcons}{\Tool{ProbCons}}
\newcommand{\Protdist}{\Tool{Protdist}}
\newcommand{\Qdist}{\Tool{QDist}}
\newcommand{\Seqgen}{\Tool{SEQ-GEN}}
\newcommand{\Teiresias}{\Tool{TEIRESIAS}}

\newcommand{\Alphabet}{\mathcal A}
\newcommand{\Vector}[1]{\mathbf #1}
\newcommand{\V}{\Vector{v}}
\newcommand{\C}{\Vector{c}}
\newcommand{\E}{\Vector{E}}
\newcommand{\Matrix}[1]{\mathbf #1}
\newcommand{\W}{\Matrix{W}}

\begin{document}

\title{Pattern-based phylogenetic distance estimation\\ and tree reconstruction}

\author{Michael H\"ohl\Address{1,2,}\footnote{Corresponding author:
Michael H\"ohl, Email: \mbox{\protect \url{m.hoehl@imb.uq.edu.au}},
Phone: +61-7-3346-2606, Fax: +61-7-3346-2101 }, Isidore
Rigoutsos\Address{2,3} and Mark A.~Ragan\Address{1,2}}

\footnotetext[1]{Institute for Molecular Bioscience, The University
of Queensland, Brisbane QLD 4072, Australia}
\footnotetext[2]{Australian Research Council Centre in Bioinformatics}
\footnotetext[3]{Bioinformatics and Pattern Discovery Group, IBM Thomas J
Watson Research Center, Yorktown Heights, NY 10598, USA}

\date{}

\maketitle

\begin{abstract}

We have developed an alignment-free method that calculates
phylogenetic distances using a maximum likelihood approach for a model
of sequence change on patterns that are discovered in unaligned
sequences.  To evaluate the phylogenetic accuracy of our method, and
to conduct a comprehensive comparison of existing alignment-free
methods (freely available as Python package \texttt{decaf+py} at
\url{http://www.bioinformatics.org.au}), we have created a dataset of
reference trees covering a wide range of phylogenetic distances.
Amino acid sequences were evolved along the trees and input to the
tested methods; from their calculated distances we infered trees whose
topologies we compared to the reference trees.

We find our pattern-based method statistically superior to all other
tested alignment-free methods on this dataset.  We also demonstrate
the general advantage of alignment-free methods over an approach based
on automated alignments when sequences violate the assumption of
collinearity.  Similarly, we compare methods on empirical data from an
existing alignment benchmark set that we used to derive reference
distances and trees.  Our pattern-based approach yields distances that
show a linear relationship to reference distances over a substantially
longer range than other alignment-free methods.  The pattern-based
approach outperforms alignment-free methods and its phylogenetic
accuracy is statistically indistinguishable from alignment-based
distances.


\end{abstract}

\section{Introduction}

Tasks like database searching, sequence classification, phylogenetic
tree reconstruction and detection of regulatory sequences are
ubiquitous in bioinformatics.  Most methods performing these tasks are
based on (automated) alignments; however, alignment-free methods exist
for solving the tasks.  Recent years have seen an increasing number of
alignment-free methods (reviewed in \citeNP{VIN:ALM:2003}, see also
\citeNP{SNE:HUY:DUT:2005}; \citeNP{VIN:GOU:ALM:2004};
\citeNP{HEL:2004}; \citeNP{PHA:ZUE:2004};
\citeNP{ULI:BUR:TUL:CHO:2005}; \citeNP{WU:HUA:LI:2005}).  In contrast
to methods based on (automated) alignments, alignment-free methods
make fewer assumptions about the nature of the sequences they work on,
and so far are mostly devoid of any evolutionary model of sequence
change (the only exception being the W-metric by
\citeNP{VIN:GOU:ALM:2004}).  The absence of an assumption of
collinearity over long stretches (implicit in any alignment) destines
them to be especially useful for handling DNA sequences that have
undergone recombination, proteins with shuffled domains, and genomic
sequences (which often feature large-scale rearrangements).

Previously, several alignment-free methods have been compared
systematically for classification purposes and their ability to detect
regulatory sequences.  However, surprisingly little is known about
their accuracy in phylogeny reconstruction.  So far, most new methods
have been verified on a few trees only.  A systematic study is sorely
lacking in this field of research.

In Section~\ref{previous-work} below we describe several
alignment-free methods that we included in our comparison.  We propose
a new alignment-free method based on patterns in sequences, and a
variant thereof, in Section~\ref{pattern-based}.  The phylogenetic
accuracy of these methods is comparatively evaluated on synthetic and
empirical data, covering a wide range of phylogenetic distances, and
we assess whether differences are statistically significant in
Section~\ref{comparison} before presenting conclusions.

\section{Previous work}
\label{previous-work}

In this section, we provide a summary of previously established
alignment-free methods.  Some methods were reviewed in
\citeA{VIN:ALM:2003}, while others are more recent and are compared
here for the first time.

We represent a biological sequence by a string $X$ of $n$ characters
taken from the alphabet $\Alphabet$ which contains $c$ different
characters $\{a_1, \dots, a_c\}$, \eg all amino acids.  Most
alignment-free methods operate on words of length $k$, so-called
\Kmers: there are $w=c^k$ such different words.  We represent the set
of \Kmers in $X$ (or a derived property) by vector $\V^X = (v^X_1,
\dots, v^X_w)$; the parameter $k$ is always implied.  Each vector
element describes the abundance of \Kmer $i$.

The (squared) Euclidean distance was introduced into sequence
comparison by \citeA{BLA:1986}. The distance between $X$ and $Y$ is
calculated using $c^X_i$, the count of \Kmer occurrences in $X$.

\begin{equation}
d^E(X,Y) = \sum_{i=1}^w \left( c^X_i - c^Y_i \right)^2
\end{equation}

Later, \citeA{BLA:1989b} found that $d^E$ yields values about twice
the number of mismatch counts obtained from alignments.

The standardized Euclidean distance was found to improve on $d^E$
without incurring the computational problems associated with the
slightly better performing Mahalanobis distance
\cite{WU:BUR:DAV:1997}.

\begin{equation}
d^S(X,Y) = \sum_{i=1}^w \left( f^X_i/s^X_i - f^Y_i/s^Y_i \right)^2
\end{equation}

Divide $f^X_i$, the relative frequencies of \Kmer occurrences in $X$,
by their standard deviations $s^X_i$ as calculated from a set of
equilibrium frequencies \cite{GEN:MUL:1989}.

\citeA{EDG:2004a} described the fractional common \Kmer count; it is
used in a distance measure that speeds up guide tree construction in
\Muscle \cite{EDG:2004b}.  Let $C^{XY}_i = \min(c^X_i, c^Y_i)$ denote
the common \Kmer count, and $Y$ be a string with $m$ characters.

\begin{equation}
F(X,Y) = \sum_{i=1}^w C^{XY}_i / [\min(n, m) - k+1]
\end{equation}

\begin{equation}
d^F(X,Y) = -\log(\epsilon+F)
\end{equation}

$F$, the fraction of common \Kmers between $X$ and $Y$, ranges from 0
to 1 and $d^F$ transforms this into a distance: $\epsilon$, a small
value added to prevent taking the logarithm of zero \cite<at least
in>{EDG:2004a}, is 0.1 there but 0.02 in \Muscle.  Both versions
employ $d^F$ in slightly different ways; here, we directly use this
common basis with $\epsilon=0.1$.

\citeA{HEL:2004} compared several metrics for their suitability in
classifying genes based on their regulatory sequences.  He found a
similarity measure based on probabilities from common \Kmer counts
under a multiplicative Poisson model to be best-performing.  In our
adaptation, we directly use the probabilities from
Equation~\ref{poissonprob} without transforming them into
similarities.

\begin{equation}\label{poissonprob}
P\negmedspace \left(x \ge C^{XY}_i \right) =
\begin{cases}
\left[ 1 - F^P\negthickspace \left( C^{XY}_i - 1, E_i \right) \right]^2
  & \text{if $C^{XY}_i > 0$,} \\
1 & \text{else.}
\end{cases}
\end{equation}

\begin{equation}
d^P(X,Y) = \biggl[ \prod_{i=1}^w P\negmedspace \left(x \ge C^{XY}_i \right) \biggr]^{1/w}
\end{equation}

Here, in the calculation of $C^{XY}_i$ the occurrence counts of \Kmers
are filtered to remove self-overlapping instances, thereby justifying
the Poisson assumption.  $F^P$ refers to the Poisson probability
distribution function and its parameter $E_i$ is the expected count
under a set of equilibrium frequencies.  $P(x \ge C^{XY}_i)$ is the
probability that we observe a \Kmer count at least as high as that
between $X$ and $Y$.

The last word-based alignment-free method considered here is the
composition distance of \citeA{HAO:QI:2004}.  Under a Markov model of
order $k-2$ we predict the probability $p^0$ of a word (the $c_i$
refer to its characters) from the probabilities of appropriate shorter
subwords, respective their relative frequencies.  To get the expected
count $E^X$ of a \Kmer in $X$, we re-arrange the corresponding total
numbers.

\begin{equation}
p^0(c_1, \dots, c_k) \equiv
\frac
{p(c_1, \dots, c_{k-1})p(c_2, \dots, c_k)}
{p(c_2, \dots, c_{k-1})}
\end{equation}

\begin{equation}
E^X(c_1, \dots, c_k) =
\frac
{c^X(c_1, \dots, c_{k-1})c^X(c_2, \dots, c_k)}
{c^X(c_2, \dots, c_{k-1})}
\cdot
\frac
{(n-k+1)(n-k+3)}
{(n-k+2)^2}
\end{equation}

We can now assemble the composition vector \cite{HAO:QI:2004} for
\Kmer occurrence counts $c^X$ in $X$: $\V^X = (\C^X - \E^X) / \E^X$.
Then we calculate the correlation between $X$ and $Y$ as the cosine of
the angle between their composition vectors, and obtain a normalized
dissimilarity $d^C$.

\begin{equation}
\cos(X,Y) =
\frac
{\sum_{i=1}^w v^X_i v^Y_i}
{\left[ \sum_{i=1}^w (v^X_i)^2 \times \sum_{i=1}^w (v^Y_i)^2 \right]^{1/2}}
\end{equation}

\begin{equation}
d^C(X,Y) = \frac{1 - \cos(X,Y)}{2}
\end{equation}

The W-metric due to \citeA{VIN:GOU:ALM:2004} is ``word-based'' but
operates on 1-mers only:

\begin{equation}
d^W(X,Y) = \sum_{i=1}^w \sum_{j=1}^w (f^X_i - f^Y_i) \cdot
(f^X_j - f^Y_j) \cdot W_{ij}
\end{equation}

Differences in amino acid composition, $f^X_i - f^Y_i$, between all
pairs of amino acids, are weighted by their entries in matrix $\W$.
\citeA{VIN:GOU:ALM:2004} found their results virtually the same for
different scoring matrices (BLOSUM62, BLOSUM50, BLOSUM40 and PAM250);
we use BLOSUM62 \cite{HEN:HEN:1992}.

\citeA{OTU:SAY:2003} showed how Lempel-Ziv complexity, computed in a
simple fashion utilizing two elementary operations
\cite{LEM:ZIV:1976}, can be used to define distance measures.  We
examine their final measure (that they call d$_1^{**}$): $c(X)$
denotes the Lempel-Ziv complexity of $X$, and $XY$ refers to the
concatenation of $X$ and $Y$.

\begin{equation}
d^{LZ}(X,Y) =
\frac
{c(XY) - c(X) + c(YX) - c(Y)}
{\frac{1}{2} \bigl[ c(XY) + c(YX) \bigr]}
\end{equation}

Most recently, \citeA{ULI:BUR:TUL:CHO:2005} proposed the Average
Common Substring (ACS) approach.  They define $L(X,Y) = \sum_{i=1}^n
\ell^{XY}_i\negmedspace/n$, where $\ell^{XY}_i$ is the length of the
longest string starting at $X_i$ that exactly matches a string
starting at $Y_j$.  $L$ provides a normalized length measure, from
which we obtain an intermediate (asymmetric) distance $d$ and finally
$d^{ACS}$.

\begin{equation}
d(X,Y) = \frac{\log(m)}{L(X,Y)} - \frac{\log(n)}{L(X,X)}
\end{equation}

\begin{equation}
d^{ACS}(X,Y) = \frac{1}{2} \bigl[ d(X,Y) + d(Y,X) \bigr]
\end{equation}

\section{Pattern-based approach}
\label{pattern-based}

We use pattern discovery to find regions of similarity (presumed
homology) occurring in two or more sequences; no alignment is
necessary.  To estimate phylogenetic distances, the patterns are
considered to be local alignments.  Adopting this point of view
enables us to apply an established maximum likelihood (ML) approach.
Both the application of pattern discovery, and distance estimation by
ML, represent novel steps in this context.  We infer pairwise
distances from the sequence data covered by patterns, yielding a
distance matrix.  Distance-based tree inference then proceeds by
conventional means.

\subsection{Terminology} 

We briefly introduce some basic terminology for \Teiresias; for more
details, including a description of the algorithm, the reader is
referred to \cite{RIG:FLO:1998a}.  Let $\Alphabet$ denote the alphabet
of characters, \eg all amino acids.  Let $'.'  \not \in \Alphabet$ be
the wildcard character that represents any amino acid.  Define a
pattern $P$ to be the regular expression $\Alphabet (\Alphabet \cup
\{'.'\})^* \Alphabet$.  A subpattern of $P$ is any substring that is a
pattern.  Call $P$ a \LW pattern ($L \leq W$) if any subpattern of
length $\geq W$ has $\geq L$ characters $\in \Alphabet$.  A pattern
$P$ has support $K$ if it occurs (has instances) in $K$ sequences.  A
pattern $P$ can be made more specific by replacing wildcard characters
by characters $\in \Alphabet$ and/or extending $P$ to the left or
right.  Call $P$ maximal \wrt a sequence set $S$ if making $P$ more
specific reduces its total support (irrespective of the number of
sequences).

We are now ready to state the behaviour of the algorithm: \Teiresias
finds all maximal \LW patterns (with support $\geq K$) in a set $S$ of
unaligned sequences.

\subsection{Distance calculation}

The pairwise distance $d^{PB}$ between two sequences $S_i$ and $S_j$
is computed as follows: first, we filter out patterns that occur more
than once in any sequence.  This removes false positives and ensures
that the self-distance of any sequence is zero.  Second, all instances
of patterns occurring simultaneously in (at least) $S_i$ and $S_j$ are
concatenated, resulting in two new sequences $S'_{ij}$ and $S'_{ji}$
of the same length.  Note that a pattern may occur in three or more
sequences, in which case we project it on multiple pairs of sequences.
Also note that generally $S'_{ij}$ (and $S'_{ji}$) will differ in
length from $S'_{ik}$ (and $S'_{ki}$) because the patterns occurring
simultaneously in (at least) $S_i$ and $S_j$ will differ in number
(and number of residues they cover) from those appearing in $S_k$ and
$S_j$.  Third, these new, concatenated sequences are used to estimate
pairwise distances.  This is done by applying a maximum
likelihood-based approach that optimizes \wrt a model of amino acid
evolution.  For the purpose of this work we use the JTT model
\cite{JON:TAY:THO:1992} as implemented in \Protdist from the \Phylip
package \cite{FEL:2005}.

Note that the algorithm for calculating distances from patterns is
general.  Our means for pattern discovery is \Teiresias, but, in
principle at least, other tools could also be used
\cite<eg>{APO:COM:PAR:2005}.  To retain the alignment-free property of
our approach, any replacement needs to have that property as well.

\subsection{Parametrization}
\label{alphabet}

Our rationale for parameterizing \Teiresias with \LWp{4}{16} is as
follows.  Consider ordinary \Kmers: higher values for $k$ reduce
chance occurrences among a set of sequences, thus reducing false
positives.  We observe that \Teiresias patterns and \Kmers bear a
relationship; to this end we introduce \emph{elementary patterns}: an
elementary pattern is a \LW pattern with exactly $L$ residues.
\Teiresias discovers maximal \LW patterns using elementary patterns as
building blocks during its convolution phase \cite{RIG:FLO:1998a}.
For the special case $W=L$ (no wildcard characters are allowed), we
find that setting $L=k$ leads to elementary patterns capturing a
subset of all \Kmers.  The only difference is that $K=1$ for \Kmers (a
\Kmer may occur only once) whereas we use $K=2$ for \Teiresias (a
pattern must occur in at least two sequences).  Thus we see that
higher values for $L$ reduce the number of false positives.  For our
distance calculation, however, we need patterns capable of accounting
for differences between sequences, hence we require $W>L$.  In
preliminary experiments on data described in Section~\ref{empirical},
we tried several higher values for $L$ with $W>L$ first.  We found for
\LWp{4}{16} (a ratio of $L/W=0.25$), the values that we use throughout
Section~\ref{comparison}, all pairwise distances are defined, \ie
every pair of sequences is covered by at least one instance of a
pattern.  For $W=8$, corresponding to a ratio of $L/W=0.5$, and higher
values of $W$, approaching the ratio $L/W=0.25$, the number of
undefined distances is 229, 127, 63, 32, 23, 8, 5, and 2 out of 8667.
(On data from Section~\ref{synthetic}, all distances are defined for
$W=8$.)

Undefined distances point towards a problem: some sequence pairs are
too divergent---no pair of substrings can be described by (elementary)
\LW patterns.  The ratio $L/W$ determines the minimum similarity any
subpattern must possess: it effectively specifies a local similarity
threshold.  Thus, undefined distances mean that no pair of substrings
reach or exceed this threshold.  Our solution to the problem is to
make sequences more similar by encoding them in a reduced alphabet.
Following \citeA{RIG:FLO:PAR:GAO:PLA:HUY:2000} and earlier work by
\citeA{TAY:1986}, we choose a reduction based on chemical
equivalences: [AG], [DE], [FY], [KR], [ILMV], [QN], [ST], [BZX] where
'[\dots]' groups similar amino acids together, and unlisted amino
acids form classes of their own.  The phylogenetic distance
calculation is based on the original sequence data covered by the
resulting patterns; this usually improves phylogenetic accuracy (see
Section~\ref{synthetic-accuracy} and~\ref{empirical-accuracy}).  As a
result of encoding sequences, all pairwise distances for \eg
\LWp{4}{8} are defined.

We also find that for sufficiently small values of $L$, the
phylogenetic accuracy is virtually independent of the particular
choice of $L$, and largely depends on the ratio $L/W$ (data not
shown).  Generally, the accuracy of tree reconstruction improves as
the local similarity threshold is lowered, with diminishing
improvement and higher computational costs the further it is lowered.

\subsection{Majority consensus and consistency}

One property of \Teiresias is that each residue can (and given our
parametrization, usually will) participate in multiple patterns.  This
may lead to situations where a particular residue in one sequence
pairs with two or more different residues in a second sequence.  It is
not clear how this should be interpreted \wrt homology.  We propose a
variant, $d^{PBMC}$ that resolves this conflict by way of (relative)
majority consensus and consistency.  We discover patterns as before
but introduce an intermediate step before distance estimation.  We
record paired positions across all patterns.  For any two sequences
$S_i$ and $S_j$, we take positions $(p_i, p_j)$ if and only if a)
$p_i$ is paired with $p'_j$ more often than with any other position in
$S_j$, b) $p_j$ is paired with $p'_i$ more often than with any other
position in $S_i$, and c) $p_i=p'_i$ and $p_j=p'_j$, \ie the positions
are the same.  This ensures that every residue participates at most
once for a given sequence pair in the distance calculation step.  For
parameters \LWp{4}{16}, the constraints prove to be stringent and
discard most of the data.

\section{Comparison of alignment-free methods}
\label{comparison}

\subsection{Synthetic data}
\label{synthetic}

We use a birth-death process to model cladogenesis
\cite{NEE:MAY:HAR:1994} and sample from several tree distributions.
The effects of different taxon sampling strategies are described in
\cite{RAN:HUE:YAN:NIE:1998}.  Trees resulting from a birth-death
process are rooted, bifurcating and ultrametric; we deviate them from
ultrametricity by an additive process to keep the expectation of the
phylogenetic distances unchanged.

Using \Phylogen V1.1 \cite{RAM:2002} we sampled seven sets of 100
four-taxon reference trees each; the parameters were
$\Opt{birth}=10.0$ and $\Opt{death}=5.0$, with $\Opt{extant} \in [40,
133, \dots, 40000]$ corresponding to a sample fraction of $[0.1, 0.03,
\dots, 0.0001]$.  The induced pairwise phylogenetic reference
distances have medians of $[0.71, 1.11, 1.61, 2.08, 2.46, 2.96, 3.39]$
substitutions per site; their upper and lower quartiles are within
$0.35$ units of these values.  For later use, we label the first,
fourth and last set as having small, medium and large phylogenetic
distances.  Sequences were evolved along the branches of the deviated
trees using \Seqgen \cite{RAM:GRA:1997} V1.3.2 under the JTT model
\cite{JON:TAY:THO:1992}, and for a sequence length of 1000 amino
acids.  (Where possible, we parameterized alignment-free methods with
the JTT model, or its equilibrium frequencies.)

To compare alignment-free methods with alignment-based methods when
the assumption of collinearity is violated, we constructed an
additional dataset with a wide distribution of phylogenetic distances.
We sampled one four-taxon tree each from 100 different distributions
specified by sample fractions that varied evenly on a logarithmic
scale.  The induced pairwise phylogenetic reference distances have a
median of 1.77 substitutions per site; the upper and lower quartiles
are 2.54 and 1.02, respectively, and the maximum is 4.88.  Sequences
of length 1000 were evolved as before, and for every sequence set the
first and last halves of two sequences were exchanged.  This
corresponds to a recent domain shuffle event.  We deliberately chose
an extreme example to show the severity that a non-justified
assumption of collinearity can have.

The generated sequences were input to the tested alignment-free
methods, and the resulting test distances were used to infer
neighbor-joining (NJ: \citeNP{SAI:NEI:1987}) trees.  Phylogenetic
accuracy is measured by the Robinson-Foulds (RF:
\citeNP{ROB:FOU:1981}) tree metric.  We compute the topological
difference between a test tree and its corresponding (unrooted)
reference tree, and report results for each set.  To assess the
statistical significance of differences between methods we employ the
Friedman test (corrected for tied ranks), followed by Tukey-style
posthoc comparisons if a significant difference is found \cite<see
e.g.>{SIE:CAS:1988}.

\subsubsection{Phylogenetic accuracy}
\label{synthetic-accuracy}

Here, and in Section~\ref{empirical-accuracy} we are interested in the
accuracy of methods in reconstructiong the phylogentic relationships
among a set of sequences; we refer to this quantity as phylogenetic
accuracy for short.  We measure and report the topological differences
between test and reference trees: better methods yield fewer
differences, and hence have a higher accuracy.  When we assess methods
based on their ranksums of the Friedman test, better methods obtain
lower numbers and rank first.

\Paragraph{Influence of k and alphabet} Here, we look at the
performance of word-based alignment-free methods as a function of the
length of \Kmers and the alphabet in use.  We varied $k$ from 1 to 9
where possible: the composition distance requires a minimum of $k=3$.
The alphabet consisted of either the original amino acids (AA) or the
chemical equivalence classes (CE) from Section~\ref{alphabet}.

For AA sequences, word length $k=4$ performs best for methods $d^E$,
$d^S$, $d^F$ and $d^P$ as judged by their ranksums based on
phylogenetic accuracy over all seven reference sets.  Second- and
third-ranking word lengths for $d^E$ and $d^P$ are $k=5$ and $k=3$.
For $d^F$ these lengths have tied ranks, and for $d^S$ this order is
reversed.  Method $d^C$ performs best for $k=3$, with $k=4$ ($k=5$)
ranking second (third).

For CE sequences, slightly higher values for $k$ yield lower ranksums.
Methods $d^E$, $d^F$ and $d^P$ perform best with word length $k=5$.
Second- and third-ranking word lengths for $d^E$ and $d^P$ are $k=6$
and $k=7$, for $d^F$ this order is reversed.  For $d^S$, word lengths
$k=5$ and $k=6$ rank equal best, followed by $k=4$.  Again, method
$d^C$ shows a preference for lower values: it performs jointly best
for $k=4$ and $k=5$, followed by $k=3$.

\begin{figure}
\centering\includegraphics[width=\textwidth]{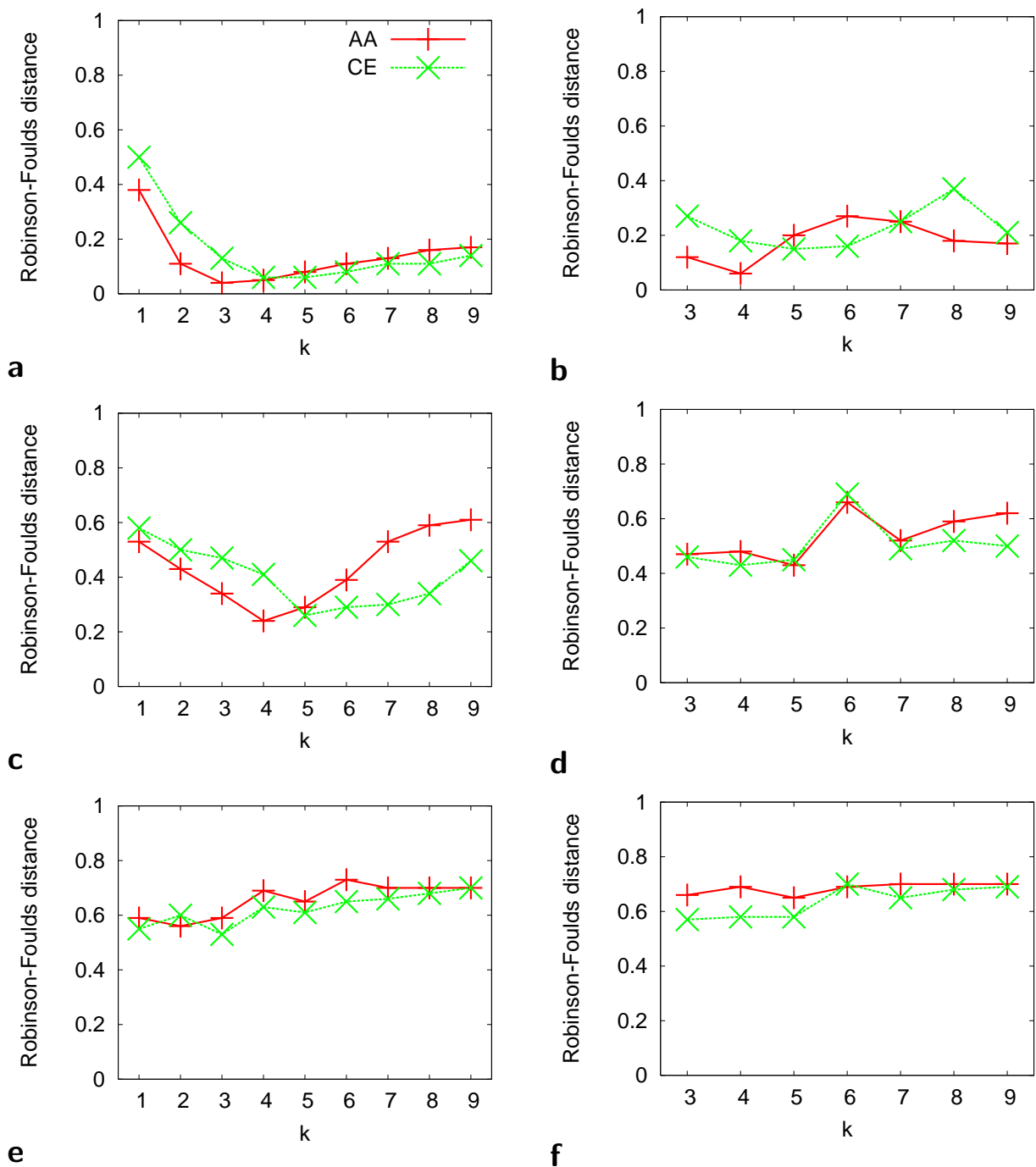}
\caption{Average Robinson-Foulds distance (Y-axis) for two methods
(a,c,e: $d^E$; b,d,f: $d^C$) on three reference sets (top to bottom:
small, medium and large phylogenetic distances).  Each subfigure shows
the behaviour as a function of $k$ (X-axis) under two alphabets (AA:
original amino acids, CE: chemical equivalence classes).  Points are
joined for ease of visual inspection only.}
\label{word-based}
\end{figure}

What we have described so far is based on the ranksums over all seven
reference sets spanning the relevant space of phylogenetic distances
for tree inference.  Looking at the phylogenetic accuracy of
word-based methods on individual sets with narrow distributions of
phylogenetic distances reveals a more complex picture.  As expected,
the topological difference between test and reference trees increases
with increasing phylogenetic reference distances.  However, depending
on the choice of $k$, the absolute values of this difference may vary
considerably.  This leads to a number of curves with different shapes
when plotting the accuracy for a particular method on the the Y-axis
with the X-axis showing values for $k$ (Figure~\ref{word-based}).  We
find that overall, the choice of method has less impact on the shape
of these curves than does phylogenetic distance.  Comparing AA with CE
sequences shows similarly shaped curves that are shifted to the right
for CE.

Figure~\ref{word-based} (a,c,e) shows curves for method $d^E$ for
three out of seven different reference sets with small, medium and
large phylogenetic distances.  The curves for methods $d^S$, $d^F$ and
$d^P$ are similar and omitted here.  The curve for medium distances
corresponds to our overall findings.  Small and large distances hint
at a better performance for small values of $k$.  Inspection of these
plots for the remaining methods fails to identify a single best $k$.
Figure~\ref{word-based} (b,d,f) for $d^C$ reveals some striking
peculiarities of this method.  In these three sets, the topological
difference for $k=6$ is often the highest, even though the neighbor
value $k=5$ may yield a low topological difference (medium distances):
thus the parameter space is uneven, more so than for other methods.

Taken together, these results indicate that, depending on the
phylogenetic distance of the sequences, the word length most
appropriate for analysis of a particular dataset may well be different
from the one performing best over all sets tested here.

\Paragraph{Statistical significance} Here, we conduct a comprehensive
comparison of all methods: we show their phylogenetic accuracy and
assign statistical significance to our findings.  For \Kmer-based
methods, we select the best-performing word lengths on the two
alphabets as described above.  We compare these methods to $d^{ACS}$,
$d^{LZ}$ and $d^W$, and to two variants of the pattern-based approach:
$d^{PB}$ and $d^{PBMC}$.  As a baseline, we include $d^{ML}$, the
maximum likelihood (ML) estimate of phylogenetic distances between the
(already correctly aligned) sequences.

Table~\ref{synthetic-incorrect} lists selected methods in ranksum
order based on all 700 trees, from best to worst.  For every method,
we show the number of incorrectly reconstructed trees in each of the
seven sets.  Recall that unrooted, bifurcating four-taxon trees can be
reconstructed either correctly or incorrectly: the RF distance will be
0 or 1, with no intermediate values possible.

\begin{table}
\begin{center}
\begin{tabular}{rrlccrrrrrrr}

 & & & & & \multicolumn{7}{c}{Synthetic reference set} \\

\\

\# & Ranksum & Method & $\Alphabet$ & $k$ & 1 & 2 & 3 & 4 & 5 & 6 & 7 \\

\\

\hline \\

 1 & 5640.0 & $d^{ML}$   & AA & --  &  2 &  2 &  7 & 12 & 13 & 18 & 17 \\

 2 & 6058.0 & $d^{PB}$   & CE & --  &  3 &  2 & 10 &  9 & 19 & 29 & 43 \\

 3 & 6390.5 & $d^{PBMC}$ & CE & --  &  3 &  3 & 10 & 10 & 23 & 42 & 59 \\

 4 & 6523.5 & $d^{PB}$   & AA & --  &  4 &  2 &  7 & 18 & 38 & 45 & 50 \\

 5 & 6951.0 & $d^S$      & CE & 5   &  8 &  6 & 19 & 27 & 45 & 47 & 57 \\

 6 & 6960.5 & $d^P$      & AA & 4   &  8 &  1 & 14 & 24 & 44 & 50 & 69 \\

 7 & 6970.0 & $d^E$      & AA & 4   &  5 &  2 & 17 & 24 & 46 & 48 & 69 \\

 8 & 6989.0 & $d^{ACS}$  & AA & --  &  9 &  3 & 17 & 24 & 49 & 52 & 59 \\

 9 & 6998.5 & $d^P$      & CE & 5   &  7 &  4 & 18 & 30 & 47 & 49 & 59 \\

10 & 7036.5 & $d^S$      & AA & 4   &  8 &  8 & 20 & 31 & 43 & 46 & 62 \\

11 & 7036.5 & $d^F$      & AA & 4   &  9 &  2 & 14 & 24 & 48 & 50 & 71 \\

12 & 7055.5 & $d^E$      & CE & 5   &  6 &  7 & 21 & 26 & 51 & 48 & 61 \\

13 & 7065.0 & $d^{ACS}$  & CE & --  & 10 &  7 & 21 & 30 & 43 & 55 & 55 \\

14 & 7074.5 & $d^{F}$    & CE & 5   &  6 &  6 & 21 & 31 & 47 & 49 & 62 \\

15 & 7359.5 & $d^{LZ}$   & CE & --  &  6 &  5 & 28 & 39 & 49 & 66 & 59 \\

16 & 7378.5 & $d^{LZ}$   & AA & --  &  3 &  8 & 21 & 34 & 53 & 64 & 71 \\

17 & 7597.0 & $d^C$      & AA & 3   & 12 & 14 & 31 & 47 & 49 & 58 & 66 \\

18 & 7635.0 & $d^C$      & CE & 5   & 15 & 14 & 32 & 45 & 54 & 63 & 58 \\

19 & 8281.0 & $d^W$      & AA & (1) & 43 & 34 & 35 & 55 & 50 & 71 & 61 \\

\end{tabular}
\end{center}
\caption{Shown are the number of incorrectly reconstructed trees (out
of 100) for each synthetic reference set and method.  The order is
based on ranksums; for each word-based method (and alphabet
$\Alphabet$), we include the best-performing $k$ (method $d^W$ can
only take on a value of 1).}
\label{synthetic-incorrect}
\end{table}

The test statistic of the Friedman test (corrected for tied ranks) is
$F_R=709.6$ ($N=700$, $k=19$).  This is highly significant
($P<10^{-10}$) beyond the $\alpha=0.05$ level.  Significant
differences are found between the following pairs (numbers refer to
column '\#' of Table~\ref{synthetic-incorrect}):  method 1 \vs methods
19--4, method 2 \vs methods 19--5, methods 3~and~4 \vs methods 19--15,
and methods 5--16 \vs method 19.  Thus the performance of most
alignment-free methods as tested here is statistically
indistinguishable from one another.  The ranksums of methods 5--14
range from 6951.0 to 7074.5, differing by $\le 123.5$.  However, the
pattern-based method $d^{PB}$ with CE, \LWp{4}{16} (ranksum: 6058.0)
is significantly better than all alignment-free methods not based on
patterns.  The ML estimate based on the correct alignment, $d^{ML}$,
is significantly better than all traditional alignment-free methods
and the pattern-based method working on original AA sequences.  Our
tests show that $d^{ML}$ is not significantly better than the two
best-performing pattern-based variants working on CE sequences.

By far the worst method tested here is the W-metric $d^W$ (ranksum:
8281.0): differences to nearly all other methods are significant.  The
lack of phylogenetic accuracy originates from being based on 1-mers.
For comparison, $d^E$ with AA, $k=1$ incorrectly reconstructs the
following number of trees for the seven reference sets: 38, 30, 39,
53, 58, 66, 59.  These numbers are quite similar to the ones in
Table~\ref{synthetic-incorrect}, as are the numbers for equally
parameterized methods $d^S$ and $d^F$.  In the case of $d^P$, however,
they are 59, 56, 65, 75, 59, 71, 65.  This is an artifact of the
method for $k=1$ (and to some extent for $k=2$) and vanishes for
higher values.  Also apparent is the poor performance of both
parametrizations of $d^{LZ}$ and $d^C$, the Lempel-Ziv and composition
distances, respectively, with ranksums between 7359.5 and 7635.0.

\Paragraph{Domain shuffling} We now describe our findings from the
reference set with simulated domain shuffling data.  We apply the same
alignment-free methods with parameter settings as before on the
unaligned, partly shuffled sequences.  Additionally, we run a number
of multiple sequence alignment (MSA) programs on this data
\cite{THO:HIG:GIB:1994,MOR:1999,EDG:2004b,DO:MAH:BRU:BAT:2005}, and
estimate ML distances from these alignments ($d^{\Clustalw}$,
$d^{\Dialign}$, $d^{\Muscle}$ and $d^{\Probcons}$, respectively).
This corresponds to an undesirable situation where \eg in an automated
environment tests have failed to detect the presence of domain
shuffling.  Hence, distances are estimated from alignments where not
all homologous residues can possibly be aligned, and it is likely that
in fact a substantial fraction of non-homologous residues have been
aligned.

The Friedman test ($F_R=270.3$, $N=100$, $k=22$) detects the presence
of a difference that is highly significant ($P<10^{-10}$) beyond the
$\alpha=0.05$ level.  However, pairwise differences are statistically
significant only between $d^{\Clustalw}$ and all other methods; this
is likely due to lack of statistical power.  Two parametrizations (CE
and AA) of the pattern-based method $d^{PB}$, \LWp{4}{16}, rank
jointly first with ranksums of 1011.5: they reconstruct 11 out of 100
trees incorrectly.  This is followed jointly by $d^{PBMC}$, with CE,
\LWp{4}{16}, and $d^S$ and $d^F$, both with CE, $k=5$.  Their ranksums
are 1066.5, and they reconstruct 16 trees incorrectly.  The numbers
for alignment-based approaches are as follows (ranksum in
parentheses):  $d^{\Clustalw}$: 71 (1671.5), $d^{\Muscle}$:  28
(1198.5), $d^{\Probcons}$: 25 (1165.5), and $d^{\Dialign}$: 21
(1121.5).  Three out of four alignment-based approaches are among the
seven worst-ranking methods.  Interestingly $d^{\Dialign}$, the
best-performing of these approaches, uses a local alignment strategy;
it occupies rank eleven jointly with two other methods.  Conversely,
what we just described means that \eg $d^{LZ}$, working on AA
sequences, one of the worst-performing alignment-free methods as
tested here, has a higher phylogenetic accuracy than three out of four
combinations of MSA program and ML estimate, and even $d^W$ is
significantly better than $d^{\Clustalw}$ on this data.  Overall, the
results show that alignment-free methods may perform better than
alignment-based approaches, especially on non-collinear sequence data,
as alignment-free methods do not make assumptions of collinearity.

\subsection{Empirical data}
\label{empirical}

We use the data from version~2 of the original \Balibase sets
\cite{THO:PLE:POC:1999a}.  They consist of 141 manually curated
benchmark alignments that are organized in five reference sets.  Their
purpose is to support tests of alignment tools under a variety of
conditions:  Set~1 is made up of roughly equi-distant sequences that
are divided into nine subsets according to their sequence conservation
and alignment length.  Set~2 contains sequence families that are
aligned with a highly divergent orphan sequence.  Set~3 aligns
subgroups with less than 25 percent identity between them.  Set~4
consists of sequences with N- or C-terminal extensions, \ie the
sequences are not trimmed at alignment boundaries.  Set~5 is
complementary to set~4:  some sequences contain internal insertions.
Two alignments contain only three sequences each and are not
considered for evaluation of phylogenetic accuracy, as there is only
one corresponding unrooted tree topology.  The remaining 139
alignments consist of between 4 and 28 sequences each.

For each reference alignment, we estimate phylogenetic reference
distances using \Protdist, and reconstruct both neighbor-joining (NJ)
and Fitch-Margoliash (FM: \citeNP{FIT:MAR:1967}) reference trees.  The
topological difference between a test tree and its corresponding
reference tree is measured by the Robinson-Foulds (RF) and the Quartet
(Q) distance (\citeNP{EST:MCM:MEA:1985}; implemented in \Qdist:
\citeNP{MAI:PED:2004}).  As these are empirical data, we cannot know
the true tree along which the sequences evolved; however, we find that
by using a large number of trees, and four combinations of tree
reconstruction method and tree topology metric (RF-NJ, RF-FM, Q-NJ,
Q-FM), we are able to rank methods robustly.  Statistical significance
is assessed as in Section~\ref{synthetic}.

\subsubsection{Phylogenetic distances}

Here, we inspect the behaviour of pairwise phylogenetic distances.
The \Balibase alignments yield 8667 reference distances, of which $<
2.5\%$ have $\ge 5.0$ substitutions per site (36 distances are $\ge
10.0$).  In what follows, we consider reference distances $< 5.0$, \ie
$< 500$ PAMs.  Note that ``distances of 250--300 PAM units are
commonly considered as the maximum for reasonable distance
estimation'' \cite{SON:HOL:2005}.  Figure~\ref{phylo-dists} contains
scatterplots where the X-axis refers to the afore-mentioned reference
distances.  The Y-axis shows the corresponding phylogenetic distances
obtained using selected alignment-free methods with original AA
sequences (and, if the methods are word-based, values for $k$ as in
Table~\ref{synthetic-incorrect}).  Additionally, we show distances
obtained from CE sequences where the distribution differs noticably.

\begin{figure}
\centering\includegraphics[width=\textwidth]{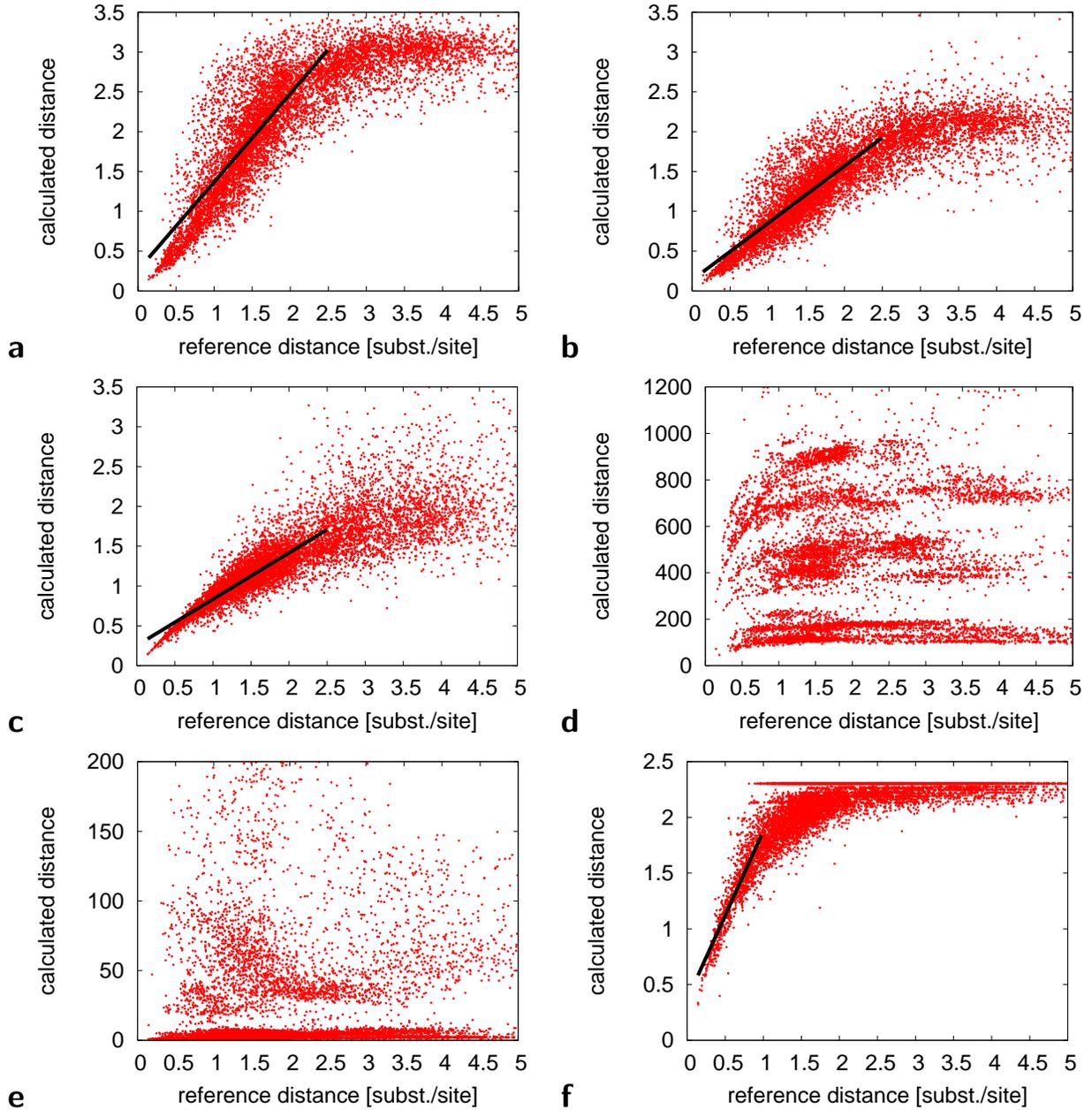}
\caption{Pairwise phylogenetic reference distances (X-axis) plotted
against corresponding calculated distances (Y-axis).  Methods and
parameters are as follows: a) $d^{PB}$ with \LWp{4}{16}, CE, b)
$d^{PB}$ with \LWp{4}{16}, AA, c) $d^{PBMC}$ with \LWp{4}{16}, CE, d)
$d^E$ with $k=4$, AA, e) $d^S$ with $k=4$, AA, f) $d^F$ with $k=4$, AA
(cont'd).}
\label{phylo-dists}
\end{figure}

\addtocounter{figure}{-1}
\begin{figure}
\centering\includegraphics[width=\textwidth]{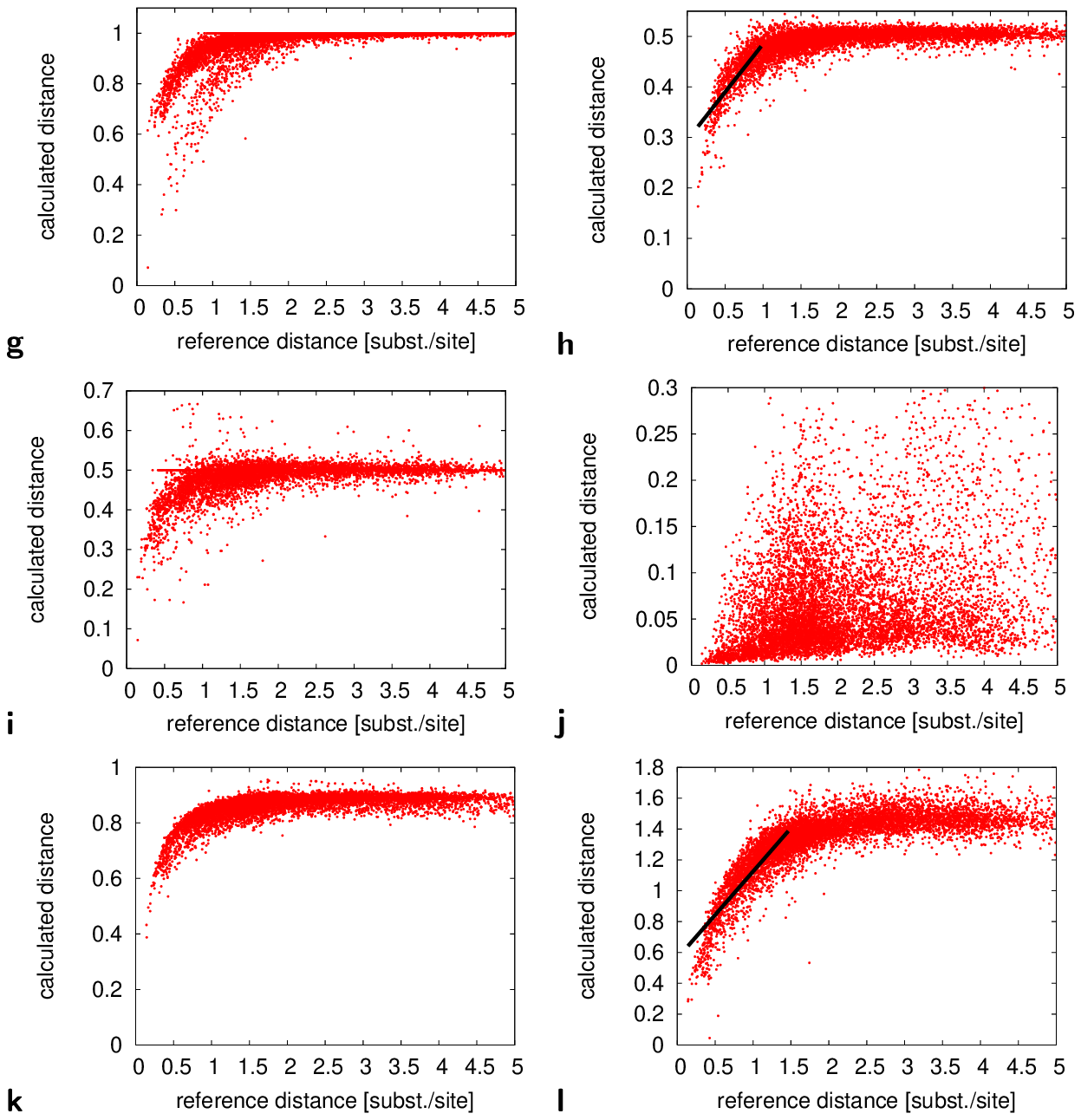}
\caption{(cont'd) g) $d^P$ with $k=4$, AA, h) $d^C$ with $k=3$, AA, i)
$d^C$ with $k=5$, CE, j) $d^{W}$, k) $d^{LZ}$ with AA, l) $d^{ACS}$
with AA.}
\end{figure}

Figure~\ref{phylo-dists}a for $d^{PB}$ with CE sequences, \LWp{4}{16},
shows a linear relationship between reference and estimated distances
for up to about 2.5 substitutions per site.  Linear regression for all
points below this cutoff yields $y=0.2598 + 1.1092x$ with a
correlation coefficient (CC) of 0.8188.  Higher distances are
increasingly underestimated as saturation comes into effect and limits
most distances ($>99\%$) to values $< 3.5$.

When patterns are discovered using the original \Balibase sequences
(Figure~\ref{phylo-dists}b) as opposed to using CE, the resulting
distances are approximated by a line with a lesser slope ($y=0.1373 +
0.7160x$, CC$=0.8261$); saturation limits most distances ($> 99\%$) to
values $< 2.5$.  For the variant $d^{PBMC}$ with CE sequences,
\LWp{4}{16} (Figure~\ref{phylo-dists}c), the linear relationship
extends at least up to 2.5 substitions per site and the slope is
roughly half that of the equally parameterized $d^{PB}$ ($y=0.2535 +
0.5819x$, CC$=0.8564$).  Again, most distances ($>99\%$) are limited
to values $< 3.5$, however this curve shows less saturation and more
scatter.

The (squared) Euclidean distance $d^E$ does not yield values that can
be interpreted in units of substitutions per site.  Instead, they
relate to mismatch counts \cite{BLA:1989b} and are therefore sequence
length-dependent.  Correspondingly, the distances do not show a single
discernable linear relationship (Figure~\ref{phylo-dists}d for $k=4$,
with AA sequences).  Most of the data ($> 99\%$) have an Euclidean
distance of $< 1200$; data for $k=5$, with CE sequences are very
similar and omitted here.

Similarly, $d^S$ has no single discernable linear relationship, with
most data ($> 99\%$) taking on numerical values of $< 200$ ($k=4$, AA,
Figure~\ref{phylo-dists}e).  However, parameters $k=5$, CE yield
distances with most values ($> 99\%$) being $< 4000$, although the
scatterplots look almost identical.

We find a linear relationship between reference and $d^F$ distances
for up to about 1.0 substitutions per site
(Figure~\ref{phylo-dists}f).  Linear regression for all points below
this cutoff yields $y=0.3681 + 1.5088x$, CC$=0.8626$.  We find 25.5\%
of all pairwise distances are $-\log(\epsilon)$, \ie would be
undefined without adding $\epsilon$ as no \Kmer is common to a
sequence pair.  This problem occurs especially for short, divergent
sequences in set~1 of \Balibase.  For $k=5$, with CE sequences, this
number drops down considerably to 15.1\%, and the plot exhibits more
scatter.

A similar problem occurs for $d^P$ (Figure~\ref{phylo-dists}g), which
is also based on common \Kmers.  Here 30.8\% of all pairwise distances
yield the value 1.0; this higher percentage is possibly caused by
numerical instabilities when multiplying small probablilities.  As
before, parametrization with $k=5$, CE reduces the percentage
considerably, to 21.0\%, and increases scatter.

The scatterplots for the composition distance $d^C$ differ between
parametrization $k=3$, AA and $k=5$, CE (Figure~\ref{phylo-dists}h,i).
The former version shows a linear behaviour for up to about 1.0
substitutions per site ($y=0.2946 + 0.1906x$, CC$=0.7881$), with no
value $\ge 0.55$, whereas the latter version shows no such large
linear behaviour and is limited to about $0.75$.  More importantly, in
this latter version 13.1\% of its distances are exactly 0.5 (within
the usual precision limits imposed by implementations of floating
point numbers), and 30.2\% are $\in [0.4999, 0.5001]$.  For the former
version, just 6 distances amount to 0.5, and 60 are $\in [0.4999,
0.5001]$.

We find no discernable relationship between reference distances $>
0.5$ and distances produced by the W-metric
(Figure~\ref{phylo-dists}j).  This likely explains why it turns out to
be the worst of all tested methods here.  The distances are mostly ($>
99\%$) limited to values $<0.3$.

The two distributions of values for parametrizations of the Lempel-Ziv
distance $d^{LZ}$ with both AA (Figure~\ref{phylo-dists}k) and CE
sequences are very similar, with CE showing more scatter.

The method $d^{ACS}$ shows a linear relationship between reference and
calculated distances for up to about 1.0--1.5 substitutions per site
(Figure~\ref{phylo-dists}l).  Linear regression for all points below
1.5 yields $y=0.5618 + 0.5631x$ (CC$=0.8643$).  We find most distances
($> 99\%$) are limited to values $< 1.65$.  For CE sequences, most
distances ($> 99\%$) are limited to values $< 1.1$.

Comparing the various distributions, we find that all three versions
of our pattern-based approach yield pairwise distances that exhibit a
linear relationship to phylogenetic reference distances for up to
about 2.5 substitutions per site.  This constitutes a considerable
increase from a maximum of 1.0--1.5 substitutions per site for methods
$d^{ACS}$, $d^F$ and $d^C$.  The linear relationship is a desirable
property, and likely explains the higher phylogenetic accuracy of the
pattern-based approach.

\subsubsection{Phylogenetic accuracy}
\label{empirical-accuracy}

\Paragraph{Statistical significance} Table~\ref{balibase-sets1to5}
lists selected alignment-free methods and four approaches based on the
ML distance estimate from automated alignments.  To obtain combined
ranksums for statistical analysis, we average the normalized
topological differences over all four combinations of tree
reconstruction method and tree distance measure (RF-NJ, RF-FM, Q-NJ,
Q-FM) based on 139 sequence sets.  The combined ranksums range from
975.5 to 2338.0.  This is a slightly wider range than for each
individual combination, for which the extreme values are $\in [987.5,
1058.5]$ and $\in [2280.0, 2312.0]$.  The first and last six methods
are ranked identically between the combined analysis and combination
RF-NJ.  The average normalized topological differences for this
combination are shown in Table~\ref{balibase-sets1to5} for all five
\Balibase reference sets.

\begin{table}
\begin{center}
\begin{tabular}{rrlccrrrrr}

 & & & & & \multicolumn{5}{c}{\Balibase reference set} \\

\\

\# & Ranksum & Method & $\Alphabet$ & $k$ & \multicolumn{1}{c}{1} & \multicolumn{1}{c}{2} & \multicolumn{1}{c}{3} & \multicolumn{1}{c}{4} & \multicolumn{1}{c}{5} \\

\\

\hline \\

 1 &  975.5 & $d^{\Muscle}$   & AA & --  & 0.240 & 0.370 & 0.274 & 0.442 & 0.244 \\

 2 & 1005.0 & $d^{\Clustalw}$ & AA & --  & 0.210 & 0.389 & 0.337 & 0.423 & 0.249 \\

 3 & 1008.0 & $d^{\Probcons}$ & AA & --  & 0.204 & 0.396 & 0.336 & 0.474 & 0.164 \\

 4 & 1190.5 & $d^{\Dialign}$  & AA & --  & 0.310 & 0.428 & 0.399 & 0.646 & 0.270 \\

 5 & 1239.0 & $d^{PB}$   & CE & --  & 0.306 & 0.510 & 0.357 & 0.478 & 0.330 \\

 6 & 1453.0 & $d^{PB}$   & AA & --  & 0.404 & 0.563 & 0.398 & 0.524 & 0.428 \\

 7 & 1570.0 & $d^{PBMC}$ & CE & --  & 0.440 & 0.557 & 0.428 & 0.609 & 0.460 \\

 8 & 1583.0 & $d^{ACS}$  & CE & --  & 0.394 & 0.583 & 0.433 & 0.591 & 0.366 \\

 9 & 1603.0 & $d^P$      & CE & 5   & 0.408 & 0.570 & 0.442 & 0.568 & 0.412 \\

10 & 1625.5 & $d^{LZ}$   & AA & --  & 0.431 & 0.569 & 0.389 & 0.642 & 0.511 \\

11 & 1632.5 & $d^E$      & CE & 5   & 0.408 & 0.593 & 0.464 & 0.570 & 0.410 \\

12 & 1646.0 & $d^F$      & CE & 5   & 0.396 & 0.575 & 0.467 & 0.569 & 0.400 \\

13 & 1703.0 & $d^{ACS}$  & AA & --  & 0.483 & 0.579 & 0.401 & 0.660 & 0.451 \\

14 & 1705.0 & $d^E$      & AA & 4   & 0.508 & 0.578 & 0.418 & 0.622 & 0.489 \\

15 & 1706.5 & $d^{LZ}$   & CE & --  & 0.421 & 0.622 & 0.440 & 0.628 & 0.437 \\

16 & 1707.5 & $d^P$      & AA & 4   & 0.496 & 0.589 & 0.419 & 0.637 & 0.469 \\

17 & 1751.5 & $d^F$      & AA & 4   & 0.515 & 0.580 & 0.431 & 0.666 & 0.475 \\

18 & 1755.0 & $d^S$      & CE & 5   & 0.446 & 0.636 & 0.491 & 0.603 & 0.375 \\

19 & 1830.0 & $d^S$      & AA & 4   & 0.513 & 0.624 & 0.450 & 0.607 & 0.528 \\

20 & 1968.5 & $d^C$      & AA & 3   & 0.481 & 0.681 & 0.525 & 0.570 & 0.588 \\

21 & 2171.0 & $d^C$      & CE & 5   & 0.535 & 0.776 & 0.642 & 0.796 & 0.611 \\

22 & 2338.0 & $d^W$      & AA & (1) & 0.585 & 0.885 & 0.795 & 0.897 & 0.720 \\

\end{tabular}
\end{center}
\caption{Shown are average topological differences for each \Balibase
reference set and method; these average values are based on
neighbor-joining trees and the normalized Robinson-Foulds measure.
The order is based on combined ranksums (see text for details);
parameters for word-based methods are as in
Table~\ref{synthetic-incorrect}.}
\label{balibase-sets1to5}
\end{table}

The Friedman test statistic $F_R=579.2$ ($N=139$, $k=22$) is highly
significant ($P<10^{-10}$) beyond the $\alpha=0.05$ level.
Significant differences are found between the following pairs (numbers
refer to column '\#' of Table~\ref{balibase-sets1to5}):  methods 1--3
\vs methods 22--6, method 4 \vs methods 22--9, method 5 \vs methods
22--12, methods 6 \vs methods 22--20, methods 7--18 \vs methods 22 and
21, and method 19 \vs method 22.  This implies that all
alignment-based approaches yield significantly better results than any
of the alignment-free methods not based on patterns, except for
$d^{\Dialign}$ \vs $d^{ACS}$ with CE.  Additionally, three out of four
alignment-based approaches (ranksums: 975.5--1008.0) are significantly
better-performing than two pattern-based variants although not than
$d^{PB}$ with CE, \LWp{4}{16} (ranksum: 1239.0).  This version
significantly outperforms all but four alignment-free methods not
based on patterns.  Again, both parametrizations of the composition
distance and the W-metric trail behind, with ranksums of 1968.5,
2171.0 and 2338.0, respectively.  Similarly to our previous analysis,
most alignment-free methods are statistically indistinguishable.
Ranksums for methods 8--18 range from 1583.0 to 1755.0, a difference
of 172.0.  On this dataset, $d^{PBMC}$ is only marginally better
(ranksum:  1570.0).  A possible explanation is apparent from
Table~\ref{synthetic-incorrect}.  There, $d^{PBMC}$ performs poorly on
reference set~7 (large phylogenetic distances) in comparison to both
parametrizations of $d^{PB}$.  We find 1986 out of 8667 pairwise
phylogenetic reference distances (\ie 22.9\%) in \Balibase have $\ge
3.0$ substitutions per site.

\section{Conclusions}

We present here for the first time a comprehensive evaluation of
alignment-free methods \wrt their accuracy in reconstructing the
phylogenetic relationship among a set of sequences.  We show that the
performance of most methods is statistically indistinguishable from
another.  The pattern-based approach as introduced by us here proved
to be significantly better than most previously established methods.
At the same time, we provide a point of reference for alignment-free
methods by measuring the maximum likelihood (ML) distance estimate
based on reference and automated alignments.  In our tests, we found
the best-performing version of our pattern-based approach $d^{PB}$ to
be statistically indistinguishable from this estimation, while most
alignment-free methods rank significantly worse on ordinary,
non-shuffled sequences.  However, on non-collinear sequences we show
that most alignment-free methods reconstruct trees more accurately
than approaches based on automated alignments.  In fact, these
alignments should not be used as they largely align non-homologous
residues.  The inspection of \Clustalw alignments reveals artifacts of
this method: it forces most residues to align with other
(non-homologous) residues, and places too few gaps.

In all three experiments we found that $d^{PB}$ ranks higher than the
equally parameterized variant $d^{PBMC}$, although not significantly.
The latter variant intuitively seems to do more justice to the concept
of homology; however, we cannot provide a satisfying explanation for
its worse perfomance.  All three versions of our pattern-based
approach result in distances that show a linear relationship to
phylogenetic reference distances over a substantially longer range
than any other alignment-free method considered here.

We also utilized a different alphabet for amino acid (AA) sequences
based on chemical equivalences (CE).  We found that $d^{PB}$ with CE
yields results as good as $d^{PB}$ with AA, and often yields
considerably increased phylogenetic accuracy.  We also tested the
other alignment-free methods on sequences encoded in this alphabet.
For any given parametrization, CE always improves performance on set~7
(large phylogenetic distances: \cf Table~\ref{synthetic-incorrect},
and also Figure~\ref{word-based} e,f) by 4 to 12 (out of 100) less
incorrectly reconstructed trees.  This probably explains why methods
parameterized with CE \vs AA perform better on \Balibase than on the
synthetic dataset.  Note that we did not try to optimize the alphabet;
certainly, there are many different choices \cite<see eg>{EDG:2004a}.
Also, our findings seem to contradict results of that study.
\citeA{EDG:2004a} found that \Kmers based on various compressed
alphabets did not improve the correlation coefficient between $d^F$
and percent identity as compared to using the original alphabet.  In
our own experiments we found the correlation coefficient between
estimated and reference distance to be a bad estimator of phylogenetic
accuracy (data not shown).

Finally, based on the data in Table~\ref{synthetic-incorrect}, we note
that there is ample room for further improvement of alignment-free
methods: compare the results for $d^{ML}$ with $d^{PB}$, especially on
reference sets 5 to 7, \ie large phylogenetic distances.  Quite likely
this will be possible only if new alignment-free methods incorporate
models of sequence change.

\section*{Acknowledgement}

M.H. received a Graduate Student Research Travel Award from the
University of Queensland.  ARC grant CE0348221 funded part of the
research.

\end{document}